\documentclass[10pt, a4paper]{article}
\usepackage{amsmath}
\usepackage{feynmp}
\usepackage{wrapfig}
\usepackage{epsfig}
\usepackage{amssymb}
\usepackage{graphicx}
\usepackage{appendix}
\usepackage{slashed}
\usepackage{indentfirst}

\makeatletter \@addtoreset{equation}{section} \makeatother

\begin{document}
\title{QUANTUM SPECTRUM OF CHERENKOV GLUE}
\author{M.N.~Alfimov$^{a,b,}$\footnote{E-mail: alfimov-mihail@rambler.ru} and A.V.~Leonidov$^{a,c,}$\footnote{E-mail: leonidov@lpi.ru} \\
$^a$~\small{\textit{P.N. Lebedev Physical Institute, Leninsky pr. 53, 119991 Moscow, Russia}} \\
$^b$~\small{\textit{Moscow Institute of Physics and Technology, Institutsky per. 9, 141900 Dolgoprudny, Russia}} \\
$^c$~\small{\textit{Institute of Theoretical and Experimental Physics, Bolshaya Cheremushkinskaya str. 25, 117218 Moscow, Russia}}
}
\date{}
\maketitle

\begin{abstract}
Full quantum calculation of Cherenkov gluon radiation by quark and gluon currents and a Cherenkov decay of a gluon into a pair of Cherenkov gluons in transparent media is performed. Energy losses due to Cherenkov gluon radiation in high energy nuclear collisions are calculated. The angular distribution of the energy flow due to the radiation of Cherenkov gluons is analyzed.
\end{abstract}

\section{Introduction}

Experimental observation of the two-humped structure of dihadron angular correlations in ultrarelativistic heavy ion collisions at RHIC \cite{STAR1,STAR2,STAR3,PHENIX1,PHENIX2,PHENIX3} bearing a remarkable likelihood to the angular distribution of Cherenkov photons \cite{cher} has brought into the focus of attention a possible existence of the phenomenon of Cherenkov radiation of gluons, an idea formulated in \cite{D79,D81} and applied to the analysis of ring-like structures in cosmic ray events in \cite{ADDK79}.

Interpretation of the experimental data in terms of the Cherenkov radiation of gluons is not unique. Theoretical descriptions aiming at describing the double-humped angular pattern of two-particle azimuthal correlations include that in terms of the Mach cone generated by jets propagating in dense medium, see e.g. the recent analysis in \cite{MW11}, as well as in terms of originating from dynamical fluctuations of the expanding hot and dense fireball \cite{SS11}. From the experimental point of view it has been demonstrated \cite{ALICE11} that in the case of large rapidity interval between the two particles and for one specific choice of transverse momenta bins for trigger and associated particles the resulting angular pattern can be completely described by the azimuthal asymmetries of the collective flow. In the case of narrow rapidity interval the situation looks different, see a detailed argumentation in \cite{MW11}, and the problem of finding an appropriate description for the experimental data at the level of detalization of \cite{STAR3} is, in our opinion, still open.

The Cherenkov radiation of gluons is a manifestation of nontrivial properties of non-Abelian medium created in ultrarelativistic heavy ion collisions \cite{DL10}. The interpretation of RHIC data in terms of Cherenkov gluon radiation and a summary of earlier work was presented in \cite{D06}. The analysis of \cite{D06} was based on a straightforward generalization of the classical Tamm-Frank theory \cite{TF}. A simple field-theoretical model of two interacting scalar fields leading to Cherenkov excitations was considered in \cite{KMW06}. A model taking into account the opacity of the medium and rescattering of Cherenkov gluons considered in \cite{DKLV09} was shown to successfully reproduce the experimental data on double-humped correlations \cite{STAR1,STAR2,STAR3,PHENIX1,PHENIX2,PHENIX3}. A new line of studies was started in \cite{CFM10a,CFM10b} where a theory of Cherenkov radiation of mesons was constructed in the framework of holographic approach to strong interactions.

To develop a more reliable theoretical picture for Cherenkov radiation of gluons one has to generalize the classical approach of \cite{D79,D81,D06} and the simple scalar field model of \cite{KMW06} to a quantum field theory description based on in-medium QCD. The main goal of the present paper is to develop such an approach to Cherenkov gluon radiation of quark and gluon currents\footnote{Some preliminary results were discussed in \cite{LA10}.}. Our consideration is essentially based on the quantum theory of electromagnetic Cherenkov radiation developed in \cite{R57}, see also \cite{STE75}. Recently the approach of \cite{STE75} was generalized to the case of a moving medium \cite{A10}.

The calculation of Cherenkov gluon radiation by quark currents presented below is a straightforward generalization of the Abelian case considered in \cite{R57}. The calculation of Cherenkov radiation of gluon currents and of the gluon decay into a pair of Cherenkov gluons are new. The corresponding expressions and the resulting qualitative picture of the pattern of energy loss related to the Cherenkov radiation present the main results of the present paper.

The plan of the paper is as follows.

In Section 2.1 we give some general remarks on the physics of Cherenkov radiation.

In Section 2.2 we compute the rate of the single Cherenkov decay of the quark current.

In Section 2.3 we compute the rate of the single Cherenkov decay of the gluon current.

In Section 2.4 we compute the rate of the double Cherenkov decay of the gluon current.

In Appendix A we describe a simple field-theoretical model justifying the Feynman rules for in-medium QCD used in the paper.

\section{Single and double Cherenkov decays}

In this section we compute the spectra of Cherenkov gluons radiated by quark and gluon currents and the spectrum of Cherenkov gluons created in the decay of a free gluon into two Cherenkov gluons.

\subsection{General remarks}

The phenomenon of Cherenkov radiation has its origin in the nontrivial changes of the dispersion relation for the excitations (quasiparticles) in the medium (as seen from the poles of the propagators):
\begin{equation}
    \frac{1}{\omega^2-{\bf k}^2} \Longrightarrow \frac{1}{\varepsilon(\omega,{\bf k})\omega^2-{\bf k}^2},
\end{equation}
where $\varepsilon(\omega,{\bf k})$ is a (chromo)permittivity of the medium under consideration.  In what follows we shall concentrate on the simplified treatment in which the standard in-vacuum quark and gluon currents interact with the transverse in-medium excitations, the Cherenkov gluons\footnote{Let us note that a more complete treatment of the problem at hand would involve a trilinear interaction of quasiparticles. A sketch of the corresponding field-theoretical formalism is given in Appendix A.}. In this setting the Cherenkov radiation is a decay of a free vacuum particle $q (g)$ into a quasiparticle ${\tilde g}$ and a free particle $q (g)$ possible for certain special values of the permittivity $\varepsilon(\omega,{\bf k}) > 1$ that allow an existence of transverse massless excitations, the Cherenkov gluons, so that, e.g., for the Cherenkov radiation of quark current we have
\begin{equation}
 q(\omega_1,{\bf k}_1) \to  q(\omega_2,{\bf k}_2) \oplus  {\tilde g} (\omega_3,{\bf k}_3).
\end{equation}
In the simplest QED case the Cherenkov radiation is a decay of a free electron into a free in-medium photon and a free electron \cite{R57}. Another interesting process to study is a decay of free in-vacuum gluons
\begin{equation}
 g(\omega_1,{\bf k}_1) \to  {\tilde g}(\omega_2,{\bf k}_2) \oplus  {\tilde g} (\omega_3,{\bf k}_3).
\end{equation}

The Cherenkov gluon emission is of course possible only for special values of energy and momenta of the three participating gluons so that the energy-momentum conservation for the considered decay is fulfilled. To give a quantitative description for this possibility one has to consider an explicit model for the chromopermittivity tensor $\varepsilon (\omega,{\bf k})$. Generically chromopermittivity is a nontrivial matrix in the color space $\varepsilon^{ab} (\omega,{\bf k})$. The nontrivial color structure of $\varepsilon^{ab} (\omega,{\bf k})$ leads, in particular, to the appearance of the color Cherenkov rainbow \cite{DL10}. In what follows we shall confine ourselves to the simplest quasi-Abelian case, where $\varepsilon^{ab} (\omega,{\bf k}) \to \delta^{ab} \varepsilon(\omega)$ and use in our qualitative estimates a  model for $\varepsilon (\omega,{\bf k})$:
\begin{align}
    & \varepsilon(\omega) = \varepsilon>1, \; \omega<\omega_0 \\
    & \varepsilon(\omega) = 1, \; \omega>\omega_0.
\end{align}
The Cherenkov radiation is then possible for excitations with energies in the interval $\omega < \omega_0$.

In what follows we shall use in our numerical estimates the values $\varepsilon = 5$ and $\omega_0=3\;{\rm GeV}$ obtained by fitting the experimental data in \cite{DKLV09}. The possible physical interpretation of $\omega_0$ is the border of the region of resonances.

\subsection{Cherenkov decay of quark current}

\begin{figure}
\begin{center}
\includegraphics[bb=75 400 545 480]{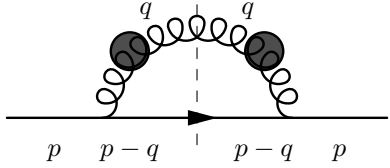}\\
\caption{Cherenkov decay of the quark current $q(p) \to q(p-q) + {\tilde g} (q)$.} \label{quark_diagram}
\end{center}
\end{figure}

Let us illustrate the approach we use in this paper by presenting a detailed calculation of the spectrum of Cherenkov gluons radiated by the massless quark current. The process in question is then a decay of a free quark into a free quark and a Cherenkov gluon,
\begin{equation}\label{qrad}
q(p) \to q(p-q) + {\tilde g} (q),
\end{equation}
where an incident quark $q(p)$ propagates along the $z$ axis and has the four-momentum $p^{\mu}=\left(E,0,0,E \right)$ and ${\tilde g}(q)$ is a Cherenkov gluon having the four-momentum $q^{\mu} =  \left(\omega,|\textbf{q}|\sin\theta,0,|\textbf{q}|\cos\theta  \right)$ and characterized by the in-medium dispersion law $\vert \mathbf{q} \vert=\sqrt{\varepsilon} \omega$ emitted at the Cherenkov angle $\theta$ with respect to the direction of the incident particle. The corresponding cut diagram is shown in Fig.~\ref{quark_diagram}\footnote{The actual calculations in the paper are performed by straightforward computation of $\vert M \vert^2$.}. The final quark has the four-momentum $p'^{\mu}=\left((E-\omega),-(E-\omega)\sin \beta,0, (E-\omega)\cos \beta \right)$.
The conservation of four-momentum in the decay leads to the following equalities fixing the Cherenkov and recoil angles $\theta$ and $\beta$:
\begin{align}
& \cos\theta = \frac{1}{\sqrt{\varepsilon}}\left(1+\frac{\varepsilon-1}{2}\frac{\omega}{E} \right) \label{cherang}, \\
& \sin \beta =  \frac{\omega}{E} \frac{1}{1-\omega/E} \sqrt{\frac{\varepsilon-1}{\varepsilon}}\left[ 1-\frac{\omega}{E}-\frac{\varepsilon-1}{4} \left(\frac{\omega}{E} \right)^2 \right]^{1/2}  \label{ptbroad}
\end{align}
The familiar classical expression for the Cherenkov angle $\cos \theta = 1/\sqrt{\varepsilon}$ follows from (\ref{cherang}) in the limit $\omega/E \to 0$. Let us note that in the energy range characterizing the trigger and associate particles, correspondingly $E$ and $\omega$, in correlation measurements in heavy ion collisions, in particular in reference to RHIC data on two-humped azimuthal angular correlations, the energy-dependent term in (\ref{cherang}) can be numerically important.

The angle $\beta$ characterizes the straggling of the incident particle in the transverse plane. From (\ref{ptbroad}) we see that in the classical limit $\omega/E \to 0$ the leading contribution to the transverse momentum of the final quark reads
\begin{align}
    & \left| \mathbf{p'}_T \right| = \omega \sqrt{\frac{\varepsilon-1}{\varepsilon}}\left[ 1-\frac{\omega}{E}-\frac{\varepsilon-1}{4} \left(\frac{\omega}{E} \right)^2 \right]^{1/2}, \\
    & \left\vert \mathbf{p'}_T \right\vert_{\omega/E \to 0} \sim \; \omega \sqrt{\frac{\varepsilon-1}{\varepsilon}} \; ,
\end{align}
so that, at given $\omega$, the relative importance of transverse struggling is directly dependent on the value of $\varepsilon$.

Let us also note that from (\ref{cherang}) there follows the restriction on the energy of Cherenkov gluon
\begin{equation}\label{radkinrestr}
    \frac{\omega}{E}\leq \frac{2}{\sqrt{\varepsilon}+1}.
\end{equation}

The matrix element for the Cherenkov decay (\ref{qrad}) $q \to q {\tilde g}$ reads
\begin{equation}
i\mathcal{M}_{q \rightarrow q {\tilde g} \;\; i \rightarrow ka}^{\quad\quad\; s \rightarrow s'j}=\overline{u}^{s'}(p-q)(-ig\gamma^{l})(t^a)_{ki}u^s(p){\tilde e}^{(j)}_{l}(q).
\end{equation}
The polarization vectors of the Cherenkov gluon $\widetilde{\textbf{e}}^{(j)}$ should satisfy the in-medium transversality condition $\textbf{q}\widetilde{\textbf{e}}^{(i)}(q)=0$ (in the present paper we use the Coulomb gauge) and can be chosen in the form\footnote{See Appendix A for details.}
\begin{equation}\label{polvec}
 \widetilde{\textbf{e}}^{(1)}(\textbf{q}) =  \frac{1}{\sqrt{\varepsilon}}(0,1,0), \;\;\;  \widetilde{\textbf{e}}^{(2)}(\textbf{q})=\frac{1}{\sqrt{\varepsilon}}(\cos\theta,0,-\sin{\theta}).
\end{equation}
Summation and averaging over the spin and color indices of the matrix element squared gives
\begin{equation}
\frac{1}{2N_c} \sum_{s,s',j,i,k,a} |\mathcal{M}_{q \rightarrow q {\tilde g} \; i \rightarrow ka}^{\quad\quad\; s \rightarrow s'j}|^2=\frac{g^2(N_c^2-1)}{N_c\varepsilon} \left(2|\textbf{p}|^2 \sin^2
\theta+2|\textbf{p}||\textbf{p}'|(1-\cos\beta) \right),
\end{equation}
from which, taking into account the dispersion law for the Cherenkov gluon, it is straightforward to compute the differential decay rate into an interval $[\omega,\omega+d\omega]$:
\begin{equation}\label{decrateqg}
\gamma_{q \rightarrow q {\tilde g}}(\omega \vert E)=\alpha_s  \frac{(N_c^2-1)}{2N_c} \left(1-\frac{1}{\varepsilon} \right) \left(1-\frac{\omega}{E}+\frac{\varepsilon+1}{4}\frac{\omega^2}{E^2} \right).
\end{equation}
As expected, it differs from the QED answer \cite{R57} only by the Casimir invariant for the fundamental representation of $SU(N_c)$,  $C_F=(N_c^2-1)/2N_c$, because we chose the simplest model of chromopermittivity, which is diagonal in the color space. The corresponding differential energy loss  per unit time is simply given by
\begin{equation}\label{difenlosqg}
P_{q \rightarrow q {\tilde g}}(\omega \vert E)= \omega \; \gamma_{q \rightarrow q {\tilde g}}(\omega \vert E).
\end{equation}
The differential energy loss spectrum (\ref{difenlosqg}) can be used for computing two observables of physical interest.

First, using the fact that there exists a one-to-one correspondence between the Cherenkov angle $\theta$ and the energy of the Cherenkov gluon $\omega$, it is straightforward to reinterpret (\ref{difenlosqg}) as describing the energy flow into an angular interval $[\theta,\theta+d\theta]$

\begin{equation}\label{}
    P(\theta)=P(\omega(\theta)|E)\frac{\partial\omega(\theta)}{\partial\theta}.
\end{equation}

\begin{figure}[ht]
\begin{minipage}[b]{0.45\linewidth}\label{f1}
\centering
   \includegraphics[height=0.23\textheight,width=\linewidth]
   {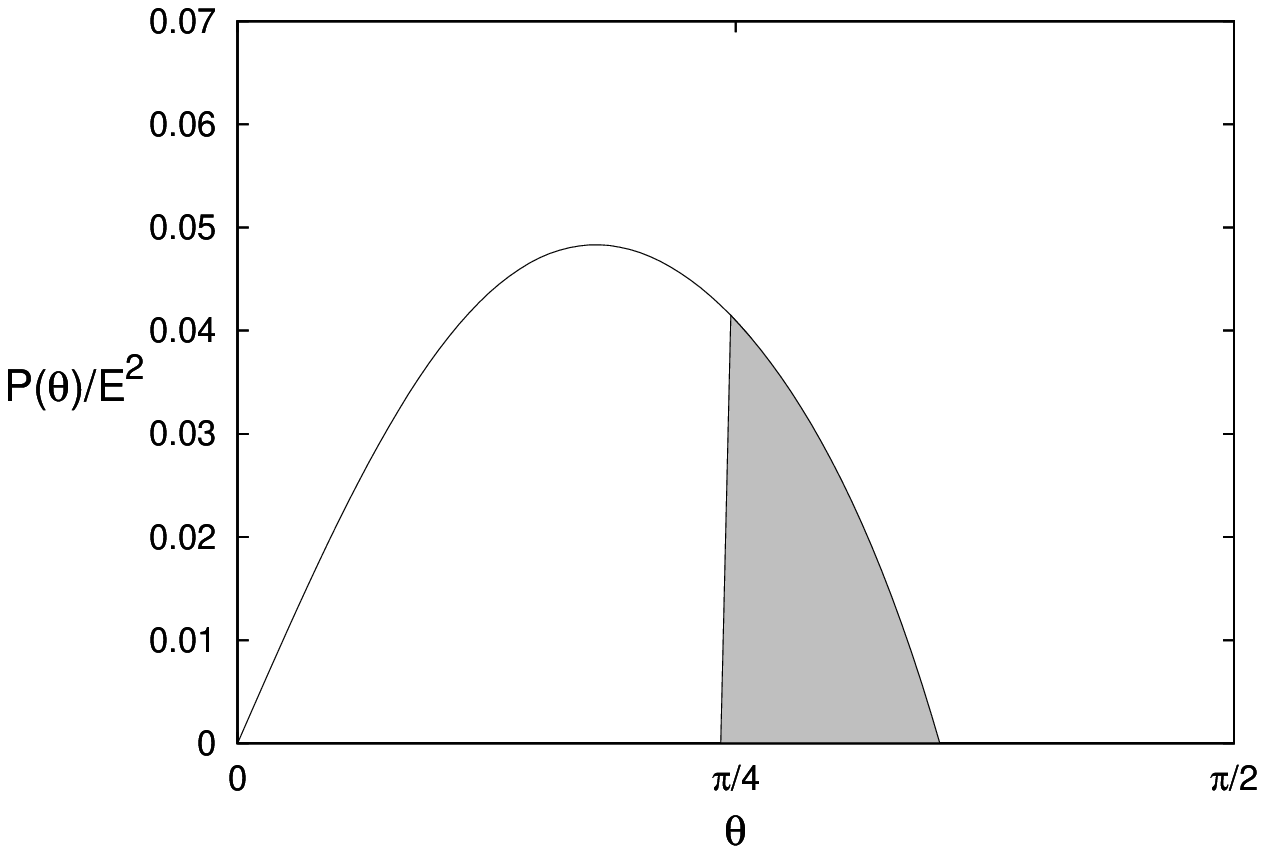}\\
   \caption{The angular differential energy flow of quark Cherenkov radiation, $\varepsilon=5$, $\omega_0=3 \; {\rm GeV}$ and $E=10 \; {\rm GeV}$.} \label{angle_quark}
\end{minipage}
\hspace{1cm}
\begin{minipage}[b]{0.45\linewidth}\label{f2}
\centering
   \includegraphics[height=0.23\textheight,width=\textwidth]{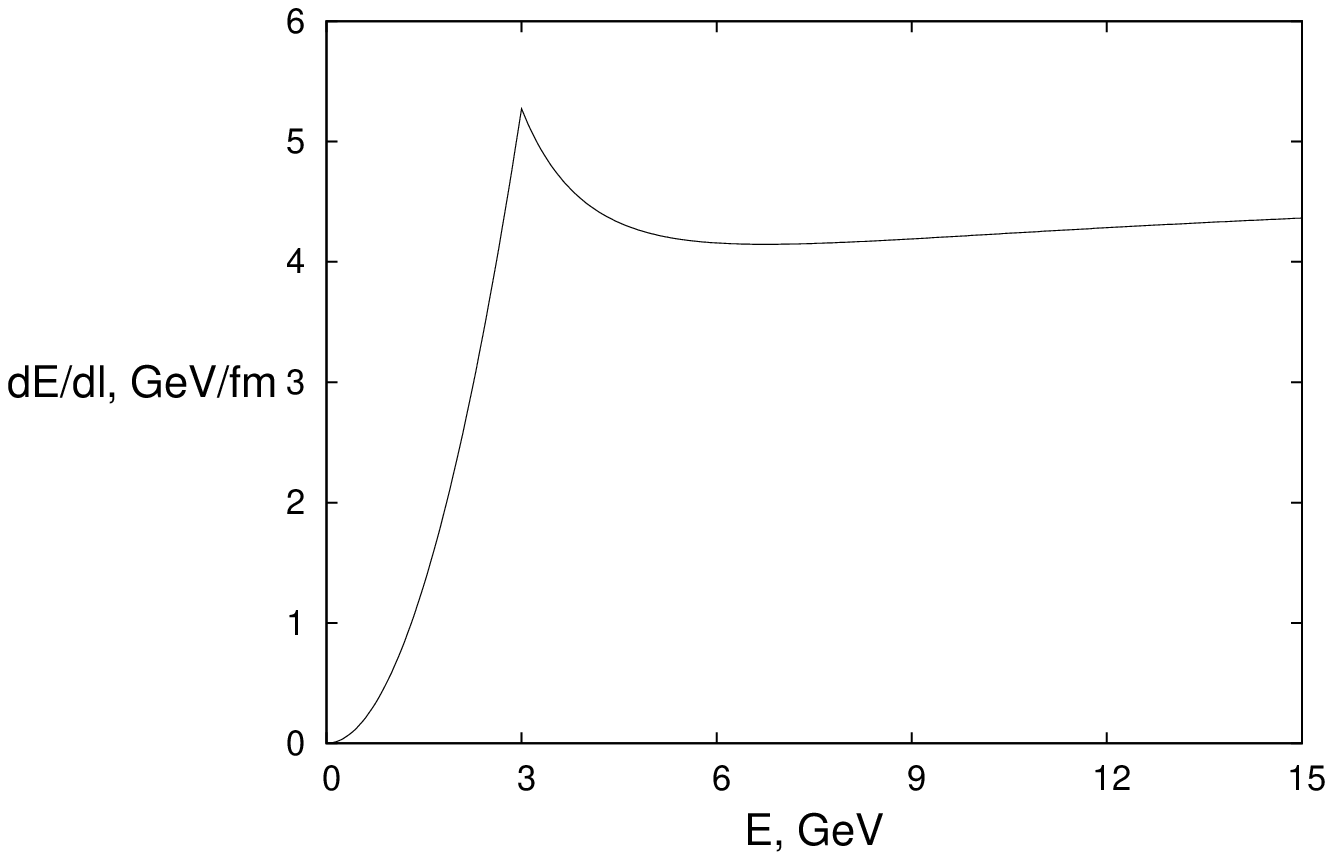}\\
   \caption{The quark Cherenkov energy loss, $\varepsilon=5$, $\omega_0=3 \; {\rm GeV}$.} \label{losses_quark}
\end{minipage}
\end{figure}

\noindent The resulting distribution is shown in Fig.~\ref{angle_quark}. We see that the energy flow is confined to an angular interval $[\theta_0,\theta_c]$ (shaded region in Fig.~\ref{angle_quark}) where the lower limit $\theta_0$ is obtained from
\begin{equation}
\cos\theta_0 = \frac{1}{\sqrt{\varepsilon}}\left(1+\frac{\varepsilon-1}{2}\frac{\omega_0}{E} \right),
\end{equation}
 and the upper limit $\theta_c$ corresponds to the classical Cherenkov angle $\cos \theta_c=1/\sqrt{\varepsilon}$ corresponding to taking the limit $\omega/E \to 0$ in (\ref{difenlosqg}).

Second, by integrating the differential spectrum (\ref{difenlosqg}) over $\omega$, one gets an expression for the energy loss per unit time:
\begin{equation}
    \frac{dE_{q \rightarrow q {\tilde g}}}{dt}(E\vert \omega_0, \varepsilon) =\int_{0}^{min\{\omega_0,\frac{2E}{\sqrt{\varepsilon}+1}\}}d\omega P_{q \rightarrow q {\tilde g}}(\omega\vert E).
\end{equation}
\noindent Note that the energy loss per unit time can be easily converted into the energy loss per unit length. The loss per unit time and per unit length are connected via the relation

\begin{equation}\label{}
    \frac{dE}{dl}=\frac{1}{v}\frac{dE}{dt},
\end{equation}

\noindent where $v$ is the speed of the incident particle. In the chosen system of units the speed of quark and gluon is $v=1$, so we have the result for the energy loss per unit length

\begin{equation}
    \frac{dE_{q \rightarrow q {\tilde g}}}{dl}(E\vert \omega_0, \varepsilon) =\int_{0}^{min\{\omega_0,\frac{2E}{\sqrt{\varepsilon}+1}\}}d\omega P_{q \rightarrow q {\tilde g}}(\omega\vert E).
\end{equation}

The resulting energy loss is plotted in Fig.~\ref{losses_quark}. We see that the Cherenkov energy loss rate for the quark current is quite substantial.

\subsection{Cherenkov decay of gluon current}

Let us now turn to the consideration of the Cherenkov gluon radiation by the gluon current. Analogously to (\ref{qrad}) this process is a decay of in-vacuum gluon into in-vacuum and Cherenkov gluons:
\begin{equation}\label{grad}
g (p) \to g(p-q) + {\tilde g} (q),
\end{equation}

\begin{figure}
\begin{center}
\includegraphics[bb=75 400 545 480]{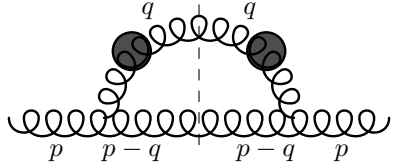}\\
\end{center}
\caption{Cherenkov decay of the gluon current $g (p) \to g(p-q) + {\tilde g} (q)$.} \label{gluon_diagram}
\end{figure}

The corresponding cut diagram for this process is shown in Fig.~\ref{gluon_diagram}. The kinematics of the gluon Cherenkov decay (\ref{grad}) is completely equivalent to that of (\ref{qrad}) and is described by the equations (\ref{cherang},\ref{ptbroad}).

The matrix element  $\mathcal{M}_{g \rightarrow g {\tilde g}}$ of the decay (\ref{grad}) reads
\begin{multline}\label{matrix_element_1}
 \mathcal{M}_{g \rightarrow g {\tilde g}}^{i \rightarrow jk}=-igf^{abc} \left[2(\textbf{p}\widetilde{\textbf{e}}^{(k)}(\textbf{q}))
 (\textbf{e}^{(i)}(\textbf{p})\textbf{e}^{(j)}(\textbf{p}'))+\right. \\
 +2(\textbf{q}\textbf{e}^{(i)}(\textbf{p}))
 (\textbf{e}^{(j)}(\textbf{p}')\widetilde{\textbf{e}}^{(k)}(\textbf{q}))- \\
 \left.-2(\textbf{q}\textbf{e}^{(j)}(\textbf{p}'))(\textbf{e}^{(i)}(\textbf{p})\widetilde{\textbf{e}}^{(k)}(\textbf{q})) \right],
\end{multline}
where the polarization vectors $\widetilde{\textbf{e}}^{(1,2)}$ are described in Eq.~(\ref{polvec}) and the conventional in-vacuum polarization vectors are

\begin{align}\label{}
    & \textbf{e}^{(1)}(\textbf{p})=\left(0,1,0 \right), \;\;\; \textbf{e}^{(2)}(\textbf{p})=\left(1,0,0 \right), \\
    & \textbf{e}^{(1)}(\textbf{p}')=\left(0,1,0 \right), \;\;\; \textbf{e}^{(2)}(\textbf{p}')=\left(\cos\beta,0,\sin\beta \right).
\end{align}

\begin{figure}
\begin{minipage}[b]{0.45\linewidth}
\centering
   \includegraphics[height=0.23\textheight,width=\linewidth]{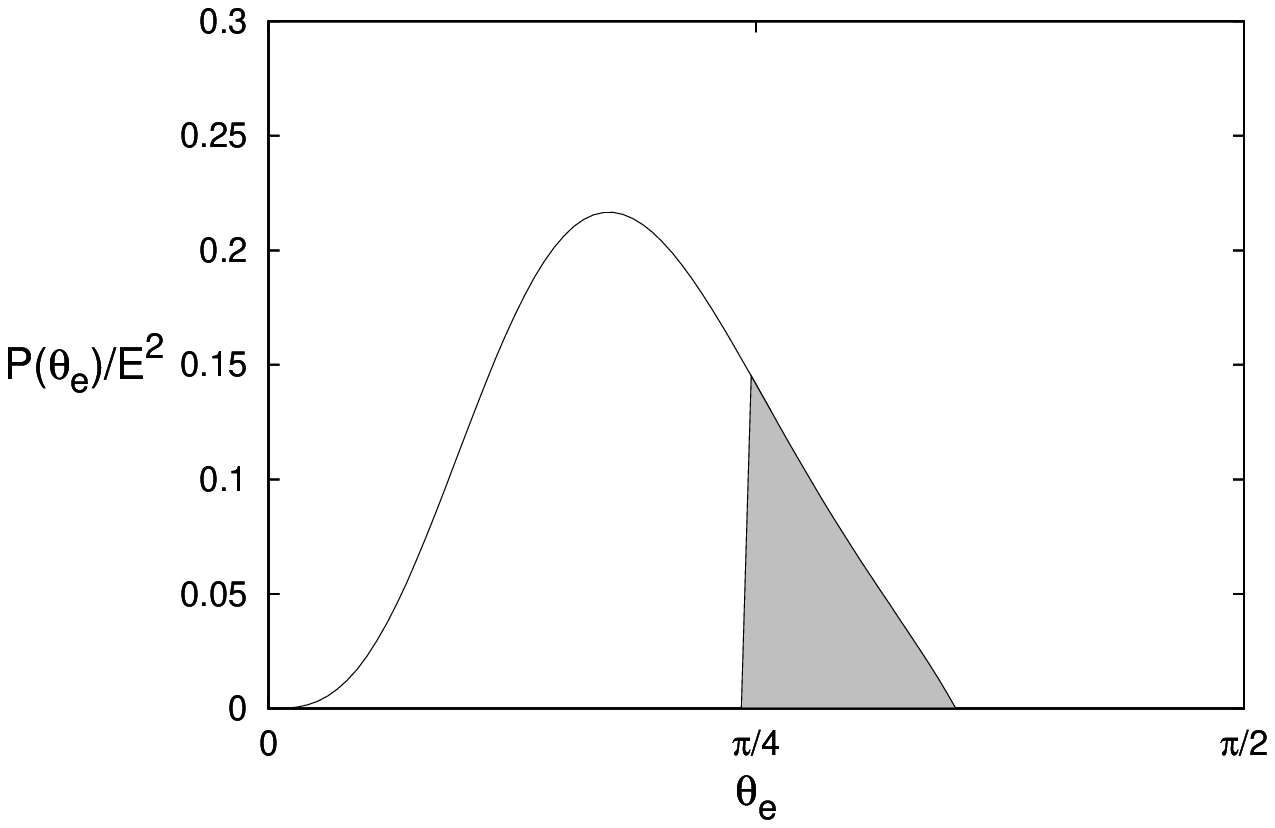}
   \caption{The angular differential energy flow of gluon Cherenkov radiation, $\varepsilon=5$, $\omega_0=3 \; {\rm GeV}$  and $E=10 \; {\rm GeV}$.} \label{angle_gluon}
\end{minipage}
\hspace{1cm}
\begin{minipage}[b]{0.45\linewidth}
\centering
   \includegraphics[height=0.23\textheight,width=\linewidth]{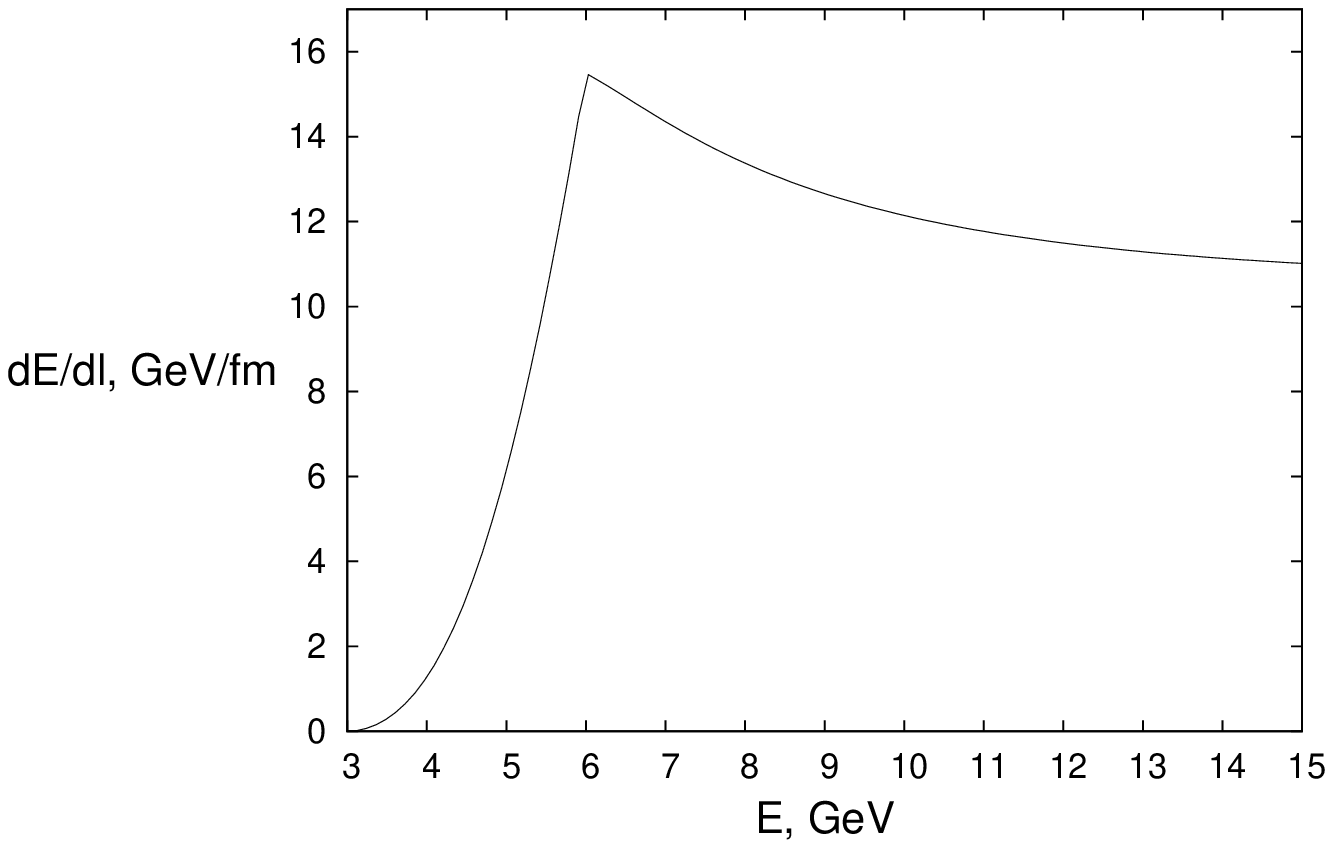}\\
   \caption{The gluon Cherenkov energy loss, $\varepsilon=5$, $\omega_0=3 \; {\rm GeV}$.} \label{losses_gluon}
\end{minipage}
\end{figure}

A straightforward computation leads to the following expression for the differential decay rate
\begin{eqnarray}\label{cherspect}
 \gamma_{g \rightarrow g {\tilde g}}(\omega \vert E) & = & \alpha_s N_c \left(1-\frac{1}{\varepsilon} \right) \left(1-\frac{\omega}{E}-\frac{\varepsilon-1}{4}
 \frac{\omega^2}{E^2} \right) \nonumber \\
& \times &
\left[
 1+\frac{1}{2}\left( \varepsilon+\frac{\varepsilon+1}{1-\frac{\omega}{E}}+\frac{\varepsilon}{\left( 1-\frac{\omega}{E} \right)^2} \right) \frac{\omega^2}{E^2}+\frac{(\varepsilon+1)^2}{8\left( 1-\frac{\omega}{E} \right)^2}\frac{\omega^4}{E^4}
\right],
\end{eqnarray}
from which one can compute in complete analogy with the calculations described in the previous paragraph. The only difference in the formula for the energy loss

\begin{equation}\label{}
    \frac{dE_{g \rightarrow g {\tilde g}}}{dl}(E\vert \omega_0, \varepsilon) =\int_{0}^{min\{\omega_0,E-\omega_0,\frac{2E}{\sqrt{\varepsilon}+1}\}}d\omega P_{g \rightarrow g {\tilde g}}(\omega\vert E),
\end{equation}

\noindent where $P_{g \rightarrow g {\tilde g}}(\omega\vert E)=\omega\gamma_{g \rightarrow g {\tilde g}}(\omega \vert E)$, is the existence of the additional restriction on the energy of the emitted Cherenkov gluon $\omega<E-\omega_0$, which is determined by the fact that the energy of the "ordinary" gluon after the emission is greater than $\omega_0$. The angular distribution of the energy flow and the rate of the energy loss shown in Figs.~\ref{angle_gluon} and ~\ref{losses_gluon} correspondingly.

The most important feature of the gluonic Cherenkov decay is the large value of the energy loss, see Fig.~\ref{losses_gluon}. This is to be expected if the cutoff energy $\omega$ is not too small and $\varepsilon$ is not too close to 1 which is definitely not the case for the values taken from the fit made in \cite{DKLV09}.

\setlength{\unitlength}{1mm}

\subsection{Double Cherenkov decay of gluon current}

\begin{figure}
\begin{center}
\includegraphics[bb=75 400 545 480]{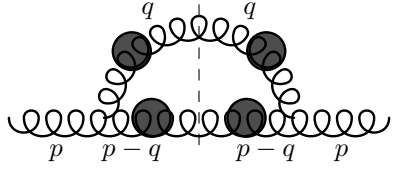}\\
\end{center}
\caption{Double Cherenkov decay of the gluon current $g (p) \to {\tilde g}(p-q) + {\tilde g} (q)$.}\label{2gluon_diagram}
\end{figure}

Let us now turn to the analysis of the another purely non-Abelian mechanism coupling ordinary gluons to the Cherenkov transverse gluonic excitations, the double Cherenkov decay
\begin{equation}\label{gdec}
g (p) \to \tilde{g}(p-q) + {\tilde g} (q),
\end{equation}
The corresponding cut diagram for this process is shown in Fig.~\ref{2gluon_diagram}.

Kinematics of the double decay (\ref{gdec}) differs from that of (\ref{qrad}) and (\ref{grad}).

First, the double decay (\ref{gdec}) is possible only for $\omega_0<E<2\omega_0$, i.e. in the restricted interval of the energy of the decaying gluon.

Second, the energy-momentum conservation laws impose restrictions on the decay angles
\begin{align}\label{decang}
    & \cos\theta=\sqrt{\varepsilon}-\frac{\varepsilon-1}{2\sqrt{\varepsilon}}\frac{E}{\omega} \\
    & \cos\beta=\sqrt{\varepsilon}-\frac{\varepsilon-1}{2\sqrt{\varepsilon}}\frac{E}{E-\omega},
\end{align}
from which we obtain the following restrictions on the energy of the emitted in-medium gluons
\begin{equation}\label{alint}
    \frac{1}{2}-\frac{1}{2\sqrt{\varepsilon}}<\frac{\omega}{E}<\frac{1}{2}+\frac{1}{2\sqrt{\varepsilon}}.
\end{equation}

Let us note that, as follows from (\ref{decang}), for the typical value $E/\omega=2$ corresponding to the center of the allowed interval (\ref{alint}) one has $\cos \theta = 1/\sqrt{\varepsilon}$, i.e. the angle equals the "classical" Cherenkov one. From the restriction that both emitted gluons must be Cherenkov gluons (with energy less than $\omega_0$) we have the restriction on their energy $E-\omega_0<\omega<\omega_0$. Then the energy flow is confined to the angular interval

\begin{equation}\label{angint}
    \sqrt{\varepsilon}-\frac{\varepsilon-1}{2\sqrt{\varepsilon}}\frac{E}{E-\omega_0} \leqslant \cos\theta \leqslant \sqrt{\varepsilon}-\frac{\varepsilon-1}{2\sqrt{\varepsilon}}\frac{E}{\omega_0}
\end{equation}

\setlength{\unitlength}{1mm}

\begin{figure}
\begin{minipage}[b]{0.45\linewidth}
\centering
   \includegraphics[height=0.23\textheight,width=\linewidth]{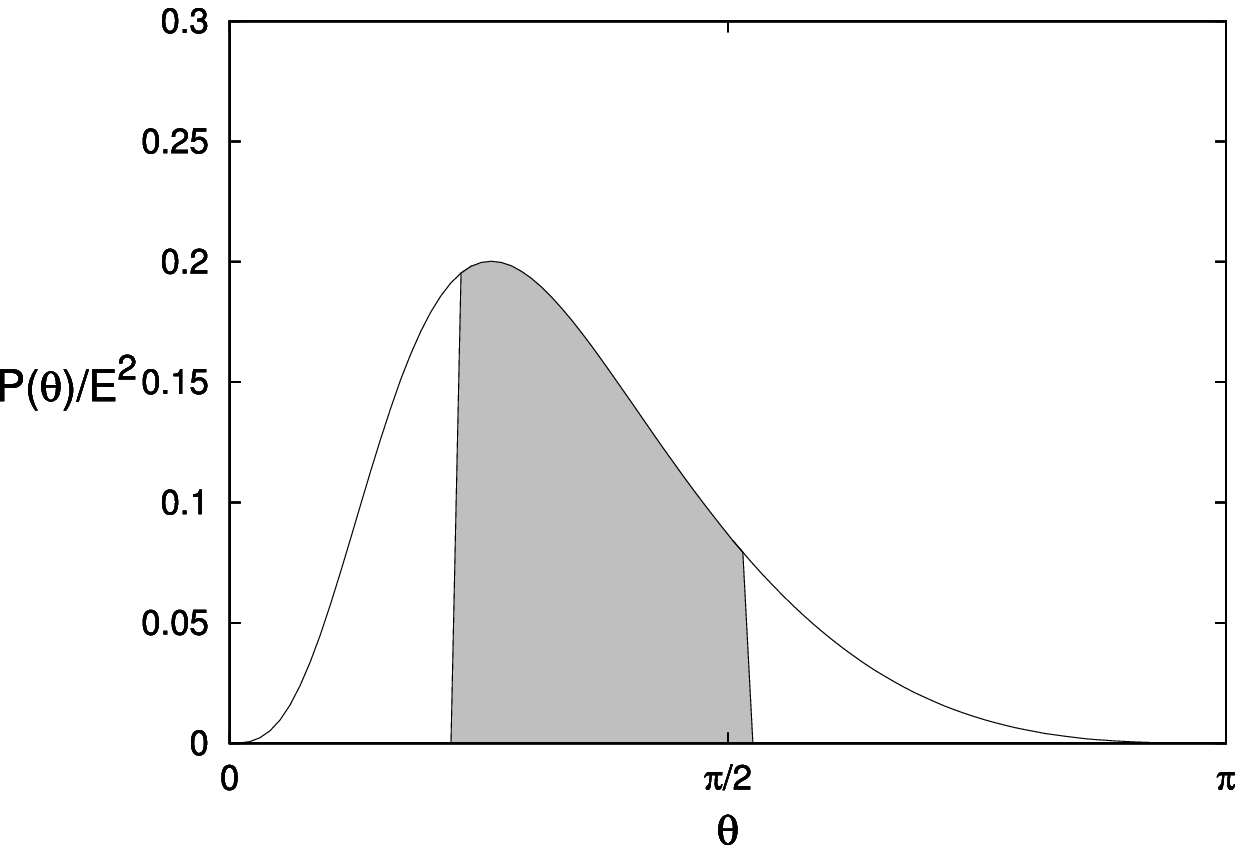}
   \caption{The angular differential energy flow of the gluonic double Cherenkov decay, $\varepsilon=5$, $\omega_0=3 \; {\rm GeV}$ and $E=5 \; {\rm GeV}$.}\label{angle_2gluon}
\end{minipage}
\hspace{1cm}
\begin{minipage}[b]{0.45\linewidth}
\centering
   \includegraphics[height=0.23\textheight,width=\linewidth]{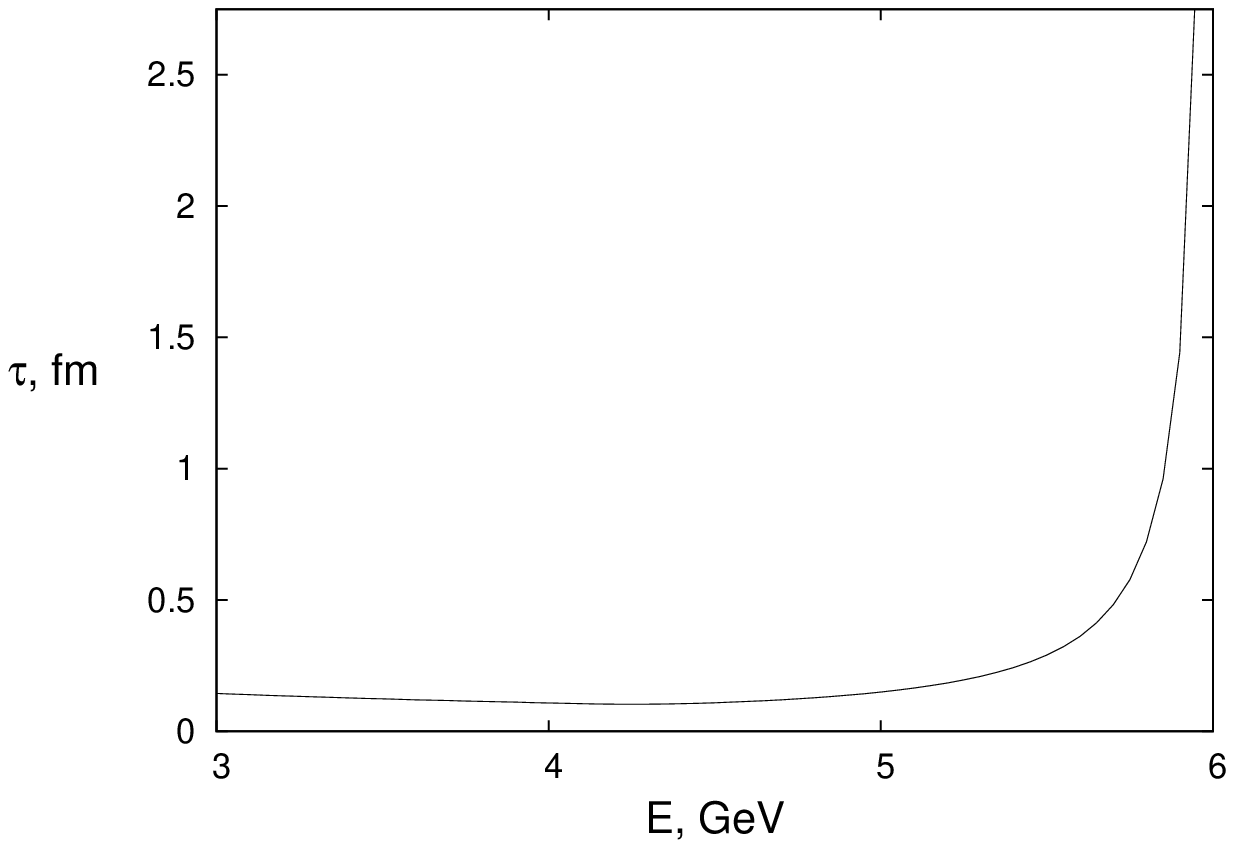}
   \caption{The lifetime of the gluon decaying through double Cherenkov decay, $\varepsilon=5$, $\omega_0=3 \; {\rm GeV}$.}\label{lifetime}
\end{minipage}
\end{figure}

The boundaries of the angular interval (\ref{angint}) tend to each other as $E$ approaches $2\omega_0$. When $E=2\omega_0$, the boundaries of the interval coincide, which physically means that the process of double Cherenkov decay have no kinematical window to take place. This conclusion is in agreement with the previous statement that the double Cherenkov decay takes place only if $\omega_0<E<2\omega_0$.

Calculation of the double Cherenkov decay is completely analogous to that for the Cherenkov decay considered in the previous paragraph. The matrix element now reads
\begin{multline}
 \mathcal{M}^{i \rightarrow jk}_{g \rightarrow {\tilde g} {\tilde g}}=-igf^{abc} \left[2(\textbf{p}\widetilde{\textbf{e}}^{(k)}(\textbf{q}))
 (\textbf{e}^{(i)}(\textbf{p})\widetilde{\textbf{e}}^{(j)}(\textbf{p}'))+\right. \\
 +2(\textbf{q}\textbf{e}^{(i)}(\textbf{p}))
 (\widetilde{\textbf{e}}^{(j)}(\textbf{p}')\widetilde{\textbf{e}}^{(k)}(\textbf{q}))- \\
 \left.-2(\textbf{q}\widetilde{\textbf{e}}^{(j)}(\textbf{p}'))(\textbf{e}^{(i)}(\textbf{p})\widetilde{\textbf{e}}^{(k)}(\textbf{q})) \right]
\end{multline}
and the corresponding expression for the differential decay rate takes the form
\begin{align}
& \gamma_{g \rightarrow {\tilde g} {\tilde g}} (\omega \vert E)= \frac{\alpha_s N_c}{2}  \left[ 1-\left(\sqrt{\varepsilon}-\frac{\varepsilon-1}{2\sqrt{\varepsilon}}\frac{E}{\omega} \right)^2 \right] \left[ 1+\varepsilon \frac{\omega^2}{E^2}+\frac{\frac{\omega^2}{E^2}}{(1-\frac{\omega}{E})^2}+\varepsilon\left( 1-\frac{\varepsilon-1}{2\varepsilon}\frac{1}{1-\frac{\omega}{E}}+\frac{\frac{\omega^2}{E^2}}{1-\frac{\omega}{E}} \right)^2 \right].
\end{align}
The angular distribution of the energy flow for the double Cherenkov decay is shown in Fig.~\ref{angle_2gluon}. The region corresponding to the allowed angular interval (\ref{angint}) is shaded. The corresponding energy loss is so large\footnote{For the initial gluon with the energy $E=5$ GeV it is about 35 GeV/fm.} (approximately three times larger than for the single Cherenkov decay considered in the previous paragraph) that it makes more sense to plot, instead of it, the lifetime of a decaying gluon as a function of its energy shown in Fig.~\ref{lifetime}. We see that unless the energy is not too close to the decay boundary of $2 \omega_0$, the decay turns out to be extremely fast.

\subsection{Cherenkov energy losses}

The results of studying the single Cherenkov decays of quark and gluon currents and the double Cherenkov decay of gluon currents lead to the following generic picture for the Cherenkov-related energy loss.
\begin{itemize}
\item{
For quark currents the only available decay channel is the single Cherenkov decay. The corresponding energy loss is non-negligible but
subleading with respect to that of the gluon current}.
\item{
For incident gluons with energy in the interval $\omega_0<E<2\omega_0$ the leading contribution to the energy loss comes from the double Cherenkov decay. The corresponding pattern of angular correlations corresponds to two peaks around the direction of propagation of the decaying gluon. There also exists a
small contribution due to single Cherenkov decay.}
\item{
At the threshold energy $E=2\omega_0$ there takes place a regime switch between the predominant double Cherenkov decay at $E<2\omega_0$ to the single Cherenkov decay of quark and gluon currents at $E>2\omega_0$ where one expects the possible appearance of the third hump corresponding to the incident particle. Besides that, as follows from (\ref{cherang}) and (\ref{decang}), at the threshold $E=2\omega_0$ there takes place the following change in the Cherenkov angle:
\begin{equation}
\cos \theta = \sqrt{\varepsilon}-\frac{\varepsilon-1}{2\sqrt{\varepsilon}}\frac{2\omega_0}{\omega} \; \to \; \cos \theta = \frac{1}{\sqrt{\varepsilon}}\left(1+\frac{\varepsilon-1}{2}\frac{\omega}{2\omega_0} \right).
\end{equation}
\noindent It is easy to verify that the Cherenkov angle of the double Cherenkov decay is greater than the Cherenkov angle for the single Cherenkov decay for all $\omega$ kinematically allowed for both processes. This means that in addition to the regime switch between the dominant energy loss processes we also have a sharp change of the Cherenkov angles of emitted gluons at $E=2\omega_0$.
}
\end{itemize}

These features appear to be in qualitative agreement with the pattern of angular correlations measured at RHIC \cite{STAR3}. A detailed comparison with the data will be published separately.

\begin{center}
{\bf Acknowledgements}
\end{center}

We are grateful to I.M. Dremin for useful discussions. Also we are grateful to D.A. Demin for help with plotting graphs.

The work of A.L. was supported by RFBR grant 09-02-00741 and CERN-RAS program. The work of M.A. was supported by the 2010 Dynasty Foundation
Grant.

\section*{Appendix}

\appendix

\section{Field theory model for in-medium QCD}

In this appendix we consider a simple field theory model justifying the Feynman rules for in-medium QCD used in the present paper. Our consideration will be confined to the case of QCD matter in its rest system.

The notion of dielectric permittivity in in-medium QED in the case of the homogeneous, isotropic medium with temporal dispersion arises in describing a response to an external electric field:
\begin{equation}
    \textbf{D}(t, \textbf{x})=\int^{t}_{-\infty} dt' \varepsilon(t-t') \textbf{E}(t', \textbf{x})=\int^{+\infty}_{0} d\tau \varepsilon(\tau) \textbf{E}(t-\tau, \textbf{x}),
\end{equation}
or, in the Fourier space:
\begin{equation}
    \textbf{D}(\omega, \textbf{r})=\varepsilon(\omega)\textbf{E}(\omega, \textbf{r}), \;\;\;\;\;\; \varepsilon(\omega)=\int^{+\infty}_{0}d\tau \varepsilon(\tau)e^{i\omega\tau}
\end{equation}

Let introduce the following simple action for the in-medium QCD (note that we use the quasi-Abelian model $\varepsilon^{ab}(t-t')=\delta^{ab} \varepsilon(t-t')$):
\begin{equation}\label{modact}
\int d^4x {\rm Tr} \left[\int^{t}_{-\infty} dt' \varepsilon(t-t') \; F^{0i}(t', \textbf{x}) \; W(t,t') \; F^{0i}(t, \textbf{x}) W^\dagger(t,t')-\frac{1}{2}F^{ij}F^{ij} \right]
\end{equation}
where
\begin{equation}
  W(t,t') = P \left\{\exp \left[ -ig\int^{t}_{t'} A^{0}(\tau, \textbf{x}) d\tau \right] \right\}
\end{equation}
is the Wilson line introduced to preserve the local gauge invariance. It is convenient to choose the Coulomb gauge $A^{a0}=0$, $\partial_{i}A^{ai}=0$ in which the action (\ref{modact}) simplifies to
\begin{equation}\label{modact1}
    S=\int d^4x \left[ \frac{1}{2}\int^{t}_{-\infty} dt' \varepsilon(t-t') \partial_0 A^{ai}(t', \textbf{x})
    \partial_0 A^{ai}(t, \textbf{x})-\frac{1}{4}F^{aij}F^{aij} \right]
\end{equation}
Let us now divide the action (\ref{modact1}) into the free field and interaction contributions:

\begin{align}\label{breakact}
    & S_0=\int d^4x \left[ \frac{1}{2}\int^{t}_{-\infty} dt' \varepsilon(t-t') \partial_0 A^{ai}(t', \textbf{x})
    \partial_0 A^{ai}(t, \textbf{x})-\frac{1}{4}(\partial^i A^{aj}-\partial^j A^{ai})(\partial^i A^{aj}-\partial^j A^{ai}) \right] \\
    & S_{int}=\int d^4x \left[ -gf^{abc}(\partial^i A^{ai}) A^{bi} A^{cj}-\frac{g^2}{4}f^{abc}f^{aed}A^{bi}A^{cj}A^{ei}A^{dj} \right]
\end{align}
From the decomposition (\ref{breakact}) there follows that in the Coulomb gauge the form of the triple-gluon interaction remains unchanged. Turning now to the free field contribution we obtain after integration by parts:
\begin{equation}\label{modact3}
    S_0=\int d^4x \frac{1}{2} \left[ \int^{t}_{-\infty} dt' \varepsilon(t-t') \partial_0 A^{ai}(t', \textbf{x})
    \partial_0 A^{ai}(t, \textbf{x})-(\partial^i A^{aj})(\partial^i A^{aj}) \right]
\end{equation}
It can be easily shown that the expression for the gluon-gluon-Cherenkov gluon and gluon-Cherenkov gluon-Cherenkov gluon vertices corresponding to the Feynman graphs is exactly the same as for the three gluon vertex in the ordinary non-Abelian gauge theory. Varying the action (\ref{modact3}) with respect to the field $A^{ai}(x)$ we arrive at the following equations of motion
\begin{equation}
    \int^{+\infty}_{0} d\tau \varepsilon(\tau) (\partial_0)^2 A^{ai}(t-\tau, \textbf{x})-(\partial_j)^2 A^{ai}(t,\textbf{x})=0
\end{equation}
or, in the Fourier space,
\begin{equation}\label{feom}
    \left(\varepsilon(k^0)(k^0)^2-\textbf{k}^2 \right)\widetilde{A}^{ai}(k)=0
\end{equation}
The above consideration shows that within the chosen model of chromoelectric permittivity we have two different branches of the dispersion relation (two different types of excitations):
\begin{align}
    & |\textbf{k}|=\sqrt{\varepsilon}\omega, \; \omega<\omega_0 \\
    & |\textbf{k}|=\omega, \; \omega>\omega_0
\end{align}
and, therefore, the following decomposition for $A^{ai}$:
\begin{align}
    & A^{ai}(x)=A_{(1)}^{ai}(x)+A_{(2)}^{ai}(x) \\
    & A_{(1)}^{ai}(x)=\int_{|\textbf{k}|<\sqrt{\varepsilon}\omega_0} \frac{d^3 k}{(2\pi)^3} \frac{1}{\sqrt{2k^0}} \sum_{\lambda} \left[ b^{a}_{\textbf{k}(\lambda)}e^{-ikx}+b^{a\dag}_{\textbf{k}(\lambda)}e^{ikx} \right] \widetilde{e}^{i}_{(\lambda)}(\textbf{k}), \; |\textbf{k}|=\sqrt{\varepsilon}k^0; \\
    & A_{(2)}^{ai}(x)=\int_{|\textbf{k}|>\omega_0} \frac{d^3 k}{(2\pi)^3} \frac{1}{\sqrt{2k^0}} \sum_{\lambda} \left[ c^{a}_{\textbf{k}(\lambda)}e^{-ikx}+c^{a\dag}_{\textbf{k}(\lambda)}e^{ikx} \right] e^{i}_{(\lambda)}(\textbf{k}), \; |\textbf{k}|=k^0;
\end{align}
It is easy to check that in order to be consistent with the commutation relations for the vector potential the
gluon polarization vectors should be normalized as follows: $(e^{i}_{(1,2)}(\textbf{k}))^2=1$ for the "ordinary" gluons and $\varepsilon(\widetilde{e}^{i}_{(1,2)}(\textbf{k}))^2=1$ for in-medium gluons.

\end{document}